# Towards stable metal inorganic-organic complex glasses

Tianzhao Xu[1], Zhencai Li[1], Kai Zheng[2], Hanmeng Zhang[3], Kenji Shinozaki[4], Huotian Zhang[5], Lars R. Jensen[6], Feng Gao[5], Jinjun Ren[3], Yanfei Zhang[2*], Yuanzheng Yue[1*]

[1]Department of Chemistry and Bioscience, Aalborg University, DK-9220 Aalborg, Denmark

[2]School of Materials Science and Engineering, Qilu University of Technology (Shandong Academy of Sciences), Jinan 250353, China

[3]Shanghai Institute of Optics and Fine Mechanics, Chinese Academy of Sciences, Shanghai 201800, China

[4]Research Institute of Core Technology for Materials Innovation, National Institute of Advanced Industrial Science and Technology (AIST), Ikeda, Osaka 563-8577, Japan

[5]Department of Physics, Chemistry, and Biology (IFM), Linköping University, Linköping 583 30, Sweden

[6]Department of Materials and Production, Aalborg University, DK-9220 Aalborg, Denmark

*Corresponding author. E-mail: zhang-yanfei@hotmail.com; yy@bio.aau.dk

**Abstract:** Metal inorganic-organic complex (MIOC) glasses have emerged as a new family of melt-quenched glasses. However, the vitrification of MIOC is challenging since most of the crystalline MIOC precursors decompose before melting. The decomposition problem severely narrows the compositional range of MIOC glass formation. Here, we report a novel approach for preparing the MIOC glasses that combines slow-solvent-removal with subsequent quenching to avoid gel thermal decomposition and crystallization. Specifically, the new approach utilizes an aprotic solvent (acetone) to kinetically prevent the ordering of the metal-ligand complex molecules in solution, thereby suppressing crystallization and forming a gel. The subsequent gradual drying process leads to the removal of the solvent to enhance the connections between molecules through hydrogen bonds, thus causing the formation of a hydrogen-bonded network. The increased network connectivity lowers the mobility of the molecules, thereby avoiding gel crystallization. Consequently, a disordered network is frozen-in during quenching of the dried gel from 130 °C to room temperature, and finally MIOC glass forms. Structural analyses reveal that hydrogen bonds are responsible for connecting the tetrahedral units. The as-prepared MIOC glass exhibits some fascinating behaviors, e.g., $T_g$ increasing with rapid room-temperature relaxation, $CO_2$ uptake, and red-shift of photoluminescence. This work not only presents a novel strategy for fabricating large-sized, stable, functional MIOC glasses, but also uncovers the critical role of hydrogen bonds in MIOC glass formation.

**Keywords:** Metal inorganic-organic complex glass; Crystallization suppression; Gel-drying-quenching; Hydrogen-bonded network; Multifunctionality

## 1. Introduction

Metal inorganic-organic complexes (MIOCs) are a subset of metal complexes (MCs). They are typically formed through the assembly of metal ions with inorganic and organic ligands into an extended network structure via coordination bonds or intermolecular interactions [1]. In 2015, a groundbreaking discovery [2] demonstrated that some crystalline MCs can undergo vitrification upon melt-quenching and are considered the new family of glass formers [3], [4], [5]. This finding sparked interest in exploring a broad family of glass-forming MCs, driven by their exceptional structural tunability and designability, which exhibited great potential for applications in gas separation [6], [7], energy storage [8], [9], and photonic [10], [11]. However, the preparation of bulk MC glasses is seriously limited by the decomposition or sublimation of organic ligands upon heating, thereby hindering the expansion of glass-forming compositions [5], [12]. To overcome this limitation, ionic liquids and organic halides have been utilized to facilitate the melting of non-meltable MCs [13], [14]. However, this strategy still cannot avoid minor decomposition and the formation of free ligands. Ali et al. established a crystallization-suppressing approach that utilizes solvents and acids to inhibit the crystallization of MCs, enabling the preparation of large-sized MIOC glasses at low temperatures and ambient conditions [15]. More recently, they further transformed MIOC glasses into amorphous MIOC foams through high-temperature treatment [16]. Despite these advances, the dependence of such approaches on pre-formed MC crystals inherently limits the flexibility in compositional tuning. Furthermore, the relatively low glass transition temperature ($T_g$ < 283 K) of the MIOC glass restricts its practical application at room temperature. The absence of a glass transition in MIOC foams indicates their amorphous rather than glassy nature, thereby leading to their reduced processability. Therefore, it is crucial to develop a straightforward and facile synthetic approach for fabricating room-temperature stable MIOC glasses with tunable composition and multi-functionalities.

In this work, we developed a crystallization-suppressing with slow-solvent-removal approach for preparing large-sized MIOC glass, i.e., $Zn(NO_3)_2(HbIm)_2$ (MIOC-$NO_3$, HbIm: protonated benzimidazole). A key advantage of this approach is that it does not involve the conventional process of both growing MC and melting crystals, thereby avoiding compositional constraints of glass formation. This study incorporated two reference materials: (1) zeolitic imidazolate framework-7 (ZIF-7, $Zn(bIm)_2$) that is a MC crystal with a coordinate bonding network, and (2) the MIOC (MIOC-Cl, $ZnCl_2(HbIm)_2$) with a hydrogen-bonded network, for comparative analysis. The glass transition behavior and phase structure were investigated using differential scanning calorimetry (DSC) and X-ray diffraction (XRD), respectively. To gain insights into the network structure, we characterized the MIOC glass using Fourier-transform infrared (FT-IR) transmission and Raman spectroscopy, high-energy synchrotron XRD, and solid-state nuclear magnetic resonance (NMR) spectroscopy. In addition, the gas absorption capacity and photonic properties of the MIOC glasses were evaluated. These comprehensive investigations not only demonstrate a novel and facile strategy for the direct preparation of large-sized MIOC glasses but also elucidate their structural features.

## 2. Experimental

### 2.1 Reagents

$Zn(NO_3)_2·6H_2O$ (Sigma, 99 %), $ZnCl_2$ (Sigma, 98 %), HbIm (Benzimidazole, Sigma, 98 %), acetone (Absolute), and ethanol (Absolute) were purchased and used for preparing the ZIF-7 crystal, MIOC-Cl crystal, and MIOC-$NO_3$ glass.

### 2.2. Preparation of MIOC crystals and glasses

MIOC-$NO_3$ glass was synthesized directly via a crystallization-suppressing method involving solution preparation, gelation, and stepwise drying. Solutions of $Zn(NO_3)_2·6H_2O$ (5 mmol) and benzimidazole (HbIm, 10 mmol) were individually dissolved in acetone (50 mL) and stirred for 30 min. The HbIm solution was then added dropwise to the zinc nitrate solution, and the mixture was stirred for another 30 min to yield a clear, transparent precursor solution with a Zn/HbIm molar ratio of 1:2. This solution was transferred to an open vial covered with perforated aluminum foil and subjected to a stepwise drying protocol in air. The solution was first dried at 378 K for 4 h, followed by ambient temperature drying for 3 days, which resulted in a light-yellow, transparent gel. The gel was further dried at 343 K for 24 h to form an orange dry gel. Finally, the dry gel was heated at 403 K for 3 days and quenched at room temperature to obtain the final transparent, orange MIOC-NO$_3$ glass.

ZIF-7 was synthesized via a solvothermal method. In a typical procedure, 10 mmol of Zn(NO$_3$)$_2$·6H$_2$O and 20 mmol of bIm were separately dissolved in 50 mL of ethanol and stirred for 15 min. The two solutions were then mixed, transferred into a stainless-steel autoclave with a Teflon lining, and heated at 373 K for 24 h. After cooling to room temperature, the resulting white powder was washed three times with ethanol and dried overnight at 343 K in air to yield the final ZIF-7 crystals.

MIOC-Cl crystal was synthesized via a slow evaporation method. 10 mmol of ZnCl$_2$ and 20 mmol of HbIm were separately dissolved in 50 mL of acetone, and each solution was stirred for 30 min. The HbIm solution was then added dropwise to the ZnCl$_2$ solution. After an additional 30 min of stirring, a clear, transparent precursor solution with a Zn/HbIm molar ratio of 1:2 was obtained. This solution was transferred to an open vial covered with perforated aluminum foil and dried at 378 K for 4 h. Subsequently, the vial was cooled to room temperature, and the remaining solution was allowed to evaporate slowly at room temperature, leading to the precipitation of white crystals. The crystals were collected, washed twice with ethanol, and dried overnight at 343 K in air to obtain the crystalline MIOC-Cl sample.

MIOC-Cl glass was prepared by the melt-quenching technique. Specifically, 100 mg of MIOC-Cl crystals were placed on a glass slide and heated to 523 K in an oven under air for 15 minutes. The melt was removed from the oven and allowed to cool rapidly in air, yielding a transparent, bulk glass product.

**2.3. Characterizations**

The crystal structure was analyzed by powder XRD on an X-ray diffractometer (Empyrean XRD, Analytical) equipped with a monochromator using Cu Kα radiation (λ = 1.5406 Å). Thermogravimetric analysis (TGA) and differential scanning calorimetry (DSC) were performed simultaneously on a Netzsch STA 449 F5 instrument under a nitrogen atmosphere with a heating rate of 10 K·min$^{-1}$. Morphology and elemental composition of the samples were characterized using a field-emission scanning electron microscopy (FE-SEM, ZEISS GeminiSEM500) coupled with an energy-dispersive X-ray spectroscopy (EDS) system.

For solution-state $^1$H NMR analysis, the MIOC-NO$_3$ glass sample (~5 mg) was digested in a mixture of 0.1 mL DCl (35 wt%) in D$_2$O and 0.5 mL deuterated dimethyl sulfoxide (DMSO-d$_6$) at 293 K. The spectra were recorded on a Bruker Avance III 500 MHz spectrometer. Chemical shifts were referenced to the residual proton signal of DMSO-$d_6$. All spectra were processed using the MestreNova Suite.

Fourier-transform infrared (FT-IR) spectra were acquired on a Bruker TENSOR II spectrometer equipped with a Bruker Platinum attenuated total reflectance (ATR) accessory over a spectral range of 400-4000 cm$^{-1}$. Raman spectra of the synthesized samples were collected on a Renishaw Via micro-Raman system with a 785 nm laser for excitation, and the spectra were recorded in the 80-2000 cm$^{-1}$ region.

Solid-state cross-polarization magic-angle spinning-nuclear magnetic resonance (CPMAS-NMR) experiments were conducted on a Bruker Avance III HD 500 MHz NMR spectrometer. For $^{13}$C-$^1$H CPMAS experiments, 3.3 μs 90° pulses and 3 ms contact time were applied on the $^1$H channel. The recycling delay was set to 8 s. Samples were spun at 6.0 kHz in a 4 mm MAS-NMR probe. The $^{13}$C-$^1$H CPMAS secondary calibration was adamantane, with the maximum intensity set to +38.5 ppm. For the $^1$H MAS experiments, a 2.3 μs 90° pulse was applied. The recycling delay was set to 8 s for the glass and 120 s for the crystal sample. Spinning was performed at 25.0 kHz using a 2.5 mm MAS-NMR probe. The $^1$H chemical shifts were referenced to the residual H$_2$O resonance at +4.8 ppm relative to tetramethylsilane (TMS, 0 ppm). For the $^{15}$N-$^1$H CPMAS experiments, a 3.3 μs 90° pulse on the $^{15}$N channel and a 3 ms contact time on the $^1$H channel were employed. The Recycle delay was 8 s for the glass and 100 s for the crystal. Spinning was conducted at 6.0 kHz in a 4 mm MAS-NMR probe. Glycine served as the secondary reference of $^{15}$N-$^1$H CPMAS, with its most intense resonance set to -347.58 ppm.

Pair Distribution Function (PDF) analysis was performed using the data of high-energy synchrotron XRD (HEXRD) measurements. The data were collected on beamline BL04B2 at SPring-8 (Hyogo, Japan) with an X-ray energy of 113 keV (Si333). For the measurements, the sample was filled into 2 mm capillaries made by silica glass. The acquired diffraction images collected by synchrotron XRD were processed using the self-made igor program [17].

Ultraviolet (UV)-Visible (VIS) absorption spectra of MIOC-NO$_3$ glass were recorded over the 200-800 nm range using a UV-VIS spectrophotometer (Cary Eclipse Fluorescence Spectrophotometer, Varian). Two-dimensional (excitation-emission) photoluminescence spectra in the 260-800 nm range were acquired using a fluorescence spectrometer (FLS1000, Edinburgh Instruments) equipped with a 450 W xenon lamp and a PMT-980 photomultiplier tube detector.

Gas adsorption-desorption isotherms for $N_2$ and $CO_2$ were measured using a Micromeritics 3Flex Surface Characterization Analyzer. The heat-treated MIOC-$NO_3$ glass was obtained by treating the as-synthesized counterpart at 523 K for 1h. Before analysis, Samples were degassed under dynamic vacuum at 373 K for 20-30 h to remove any adsorbed species. $N_2$ adsorption isotherms were recorded at 77 K, while $CO_2$ isotherms were measured at 195 K.

## 3. Results and discussion

### 3.1. Characteristics of MIOC glass

The orange, transparent bulk glass of MIOC-$NO_3$ [$Zn(NO_3)_2(HbIm)_2$] was directly synthesized using a crystallization-suppressing method involving a slow-solvent-removal process. The preparation procedure is illustrated in Figure 1a. Figure 1b shows photographs of the as-prepared MIOC-$NO_3$ glass and its remelted bulk glass from ground MIOC-$NO_3$ glass powder, which illustrates the excellent processability of it. For structural comparison, ZIF-7 crystals were synthesized via a solvothermal method, while MIOC-Cl crystals were prepared via a slow evaporation method. The corresponding MIOC-Cl glass was prepared by melt-quenching its crystalline counterpart (see Method part). A key structural similarity among these materials is the presence of Zn metal nodes and bIm (or HbIm) ligands. The XRD patterns of all synthesized crystals and glasses are presented in Figure 1c. The diffraction patterns of ZIF-7 and MIOC-Cl can be indexed to monoclinic [18] and triclinic phases [19], respectively. Conversely, both the gel-drying-quenched MIOC-$NO_3$ glass and the melt-quenched MIOC-Cl glass exhibit a broad halo pattern without any sharp Bragg peaks, indicating their amorphous nature. The cross-sectional SEM image of the MIOC-$NO_3$ glass (Figure S1) displays a featureless and smooth morphology, and the corresponding elemental mapping confirms a homogeneous distribution of all constituent elements. To verify the complete removal of the solvent, the as-prepared glass was dissolved in DCl and analyzed by liquid-state $^1$H NMR (Figure S2). The $^1$H NMR spectrum is dominated by signals from acetone [20] and HbIm ligands [21]. Quantitative integration indicates a minimal residual acetone content of only 0.6 mol%, confirming the high efficiency of the solvent removal process. The presence of additional faint signals in the spectrum suggests minor ligand decomposition during the preparation process, which is likely responsible for the observed orange color of the glass. The negligible amount of residual solvent confirms that the amorphous structure of the MIOC-$NO_3$ glass is an intrinsic feature, rather than the solvent-induced structural disorder.

The DSC curve of the MIOC-$NO_3$ glass (Figure 1d) reveals a distinct glass transition starting at 348 K ($T_g$), followed by decomposition at 454 K ($T_d$). The absence of a desolvation peak prior to decomposition confirms the effective removal of the solvent. Furthermore, a wide processing window of 106 K ($T_d$ - $T_g$) highlights its good thermal stability for further processing. A notable phenomenon could be observed in the consecutive DSC heating cycles (Figure 1e) that the $T_g$ of the as-prepared glass decreased from 348 K to a stable value of 333 K after the first scan. This shift in $T_g$, occurring without any accompanying mass change, is characteristic of the sub-$T_g$ relaxation of glass. That is, during storage at room temperature, the glass undergoes structural relaxation, a process that reduces its free volume. This relaxation subsequently manifests as an endothermic overshoot in the DSC scan, leading to an increase in the apparent $T_g$ [22], [23]. For comparison, the thermal behaviors of the ZIF-7 and the MIOC-Cl are examined (Figures S3 and S4). ZIF-7 is thermally stable with no transitions before decomposition at 772 K. In contrast, MIOC-Cl melts at 519 K and can be quenched into a glassy state with a $T_g$ of 361 K. The close $T_g$ values of the MIOC-$NO_3$ (348 K) and MIOC-Cl (361 K) suggest that they might share a similar underlying network structure, likely dominated by hydrogen bonding. Both the higher thermal stability of ZIF-7 and the absence of melting event imply a significantly robust coordination bond network in ZIF-7 that inhibits a structural rearrangement needed for decomposition and melting.

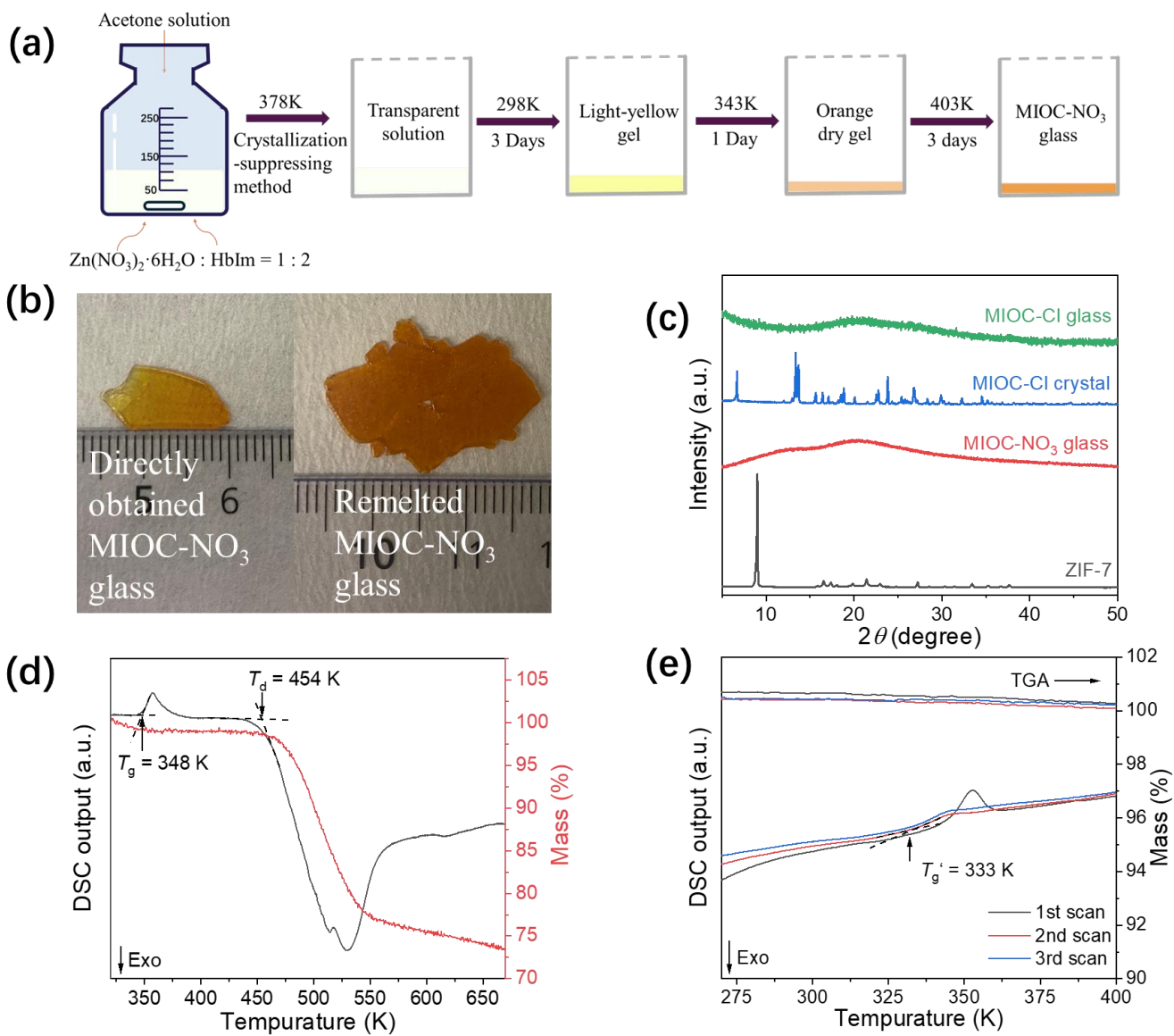

**Figure 1. (a)** Schematic diagram of the synthesis process of MIOC-NO$_3$ glass. **(b)** Photographs of both directly obtained and remelted MIOC-NO$_3$ glasses. **(c)** XRD patterns of the synthesized MIOC-NO$_3$ glass, ZIF-7 crystals, MIOC-Cl crystal, and MIOC-Cl glass. **(d)** DSC and Thermogravimetric analysis (TGA) curves of the as-synthesized MIOC-NO$_3$ glass. **(e)** DSC-TGA curves of MIOC-NO$_3$ glass obtained by three consecutive scans.

### 3.2. Structural analysis

The bonding modes in the network structure of MIOC-NO$_3$ glass, along with the other three reference samples, were investigated by the FT-IR and Raman spectra, respectively (Figure 2). As shown in Figure 2a, the FT-IR spectra confirm the presence of the HbIm/bIm ligands in all materials, as evidenced by peaks at 434 and 422 cm$^{-1}$, which correspond to the in-plane ring bending mode of HbIm/bIm [24]. It is noted that the FT-IR spectra provide critical insight into the chemical state of the nitrate anions in the MIOC-$NO_3$ glass. The presence of a peak at 811 cm$^{-1}$, assigned to the N-O bending mode bonds [25] and a peak at 1477 cm$^{-1}$, attributed to the stretching mode of $C_{2v}$ symmetric coordinated $NO_3$ ligands, indicates that nitrate groups are bound within the network. However, the characteristic peak of $D_{3h}$ symmetric free nitrate ion at 1370 cm$^{-1}$ is completely absent, definitively confirming that the nitrate groups are fully coordinated within the glass network, rather than existing as free ions [26]. In contrast, the peak corresponding to the stretching mode of the Zn-Cl bond is located at 300 cm$^{-1}$, which is beyond the detection range, and no further information can be obtained [27]. Apart from the $NO_3$-related features, the consistency of the FT-IR spectra of MIOC-Cl and MIOC-$NO_3$ glasses suggests that they share a similar metal-ligand correlation.

The most distinct difference is observed in the hydrogen-bonding region from 3300 to 2900 cm$^{-1}$. Whereas ZIF-7 shows no absorption in this region, all MIOC-based samples exhibit characteristic N-H stretching vibrations. Specifically, the peaks located at approximately 3320 cm$^{-1}$ (MIOC-Cl crystals), 3204 cm$^{-1}$ (MIOC-Cl glass), and 3145 cm$^{-1}$ (MIOC-$NO_3$ glass) are attributed to the stretching modes of N-H bonds. The progressive broadening and red shift of the N-H bond peaks are a direct spectroscopic signature of the increasing hydrogen bond strength [28]. The role of hydrogen bonding in facilitating glass formation is clarified by a comparison of these three distinct Zn-based frameworks. In MIOC-Cl crystals, the N-H⋯Cl hydrogen bonds, formed by N-H bonds in the HbIm ligands and Cl atoms on adjacent Zn tetrahedra, are weakened by steric hindrance and other intermolecular interactions. However, upon vitrification, a reorganization of these interactions occurs, establishing a disordered hydrogen bond network dominated by N-H⋯Cl bonds. The MIOC-$NO_3$ glass features strong N-H⋯O hydrogen bonds between the bIm ligands and $NO_3$ ligands. This interaction enables the direct formation of a disordered hydrogen bond network from the solution state, suppressing the crystallization of MIOC. In contrast, the ZIF-7 crystal, which lacks hydrogen bonding and depends solely on a rigid coordination bond network, does not form a glass. This comparative analysis demonstrates that hydrogen bonding is a critical factor for inducing the structural disorder necessary for MIOC glass formation, as it promotes the formation of a flexible, disordered network over a rigid crystalline one.

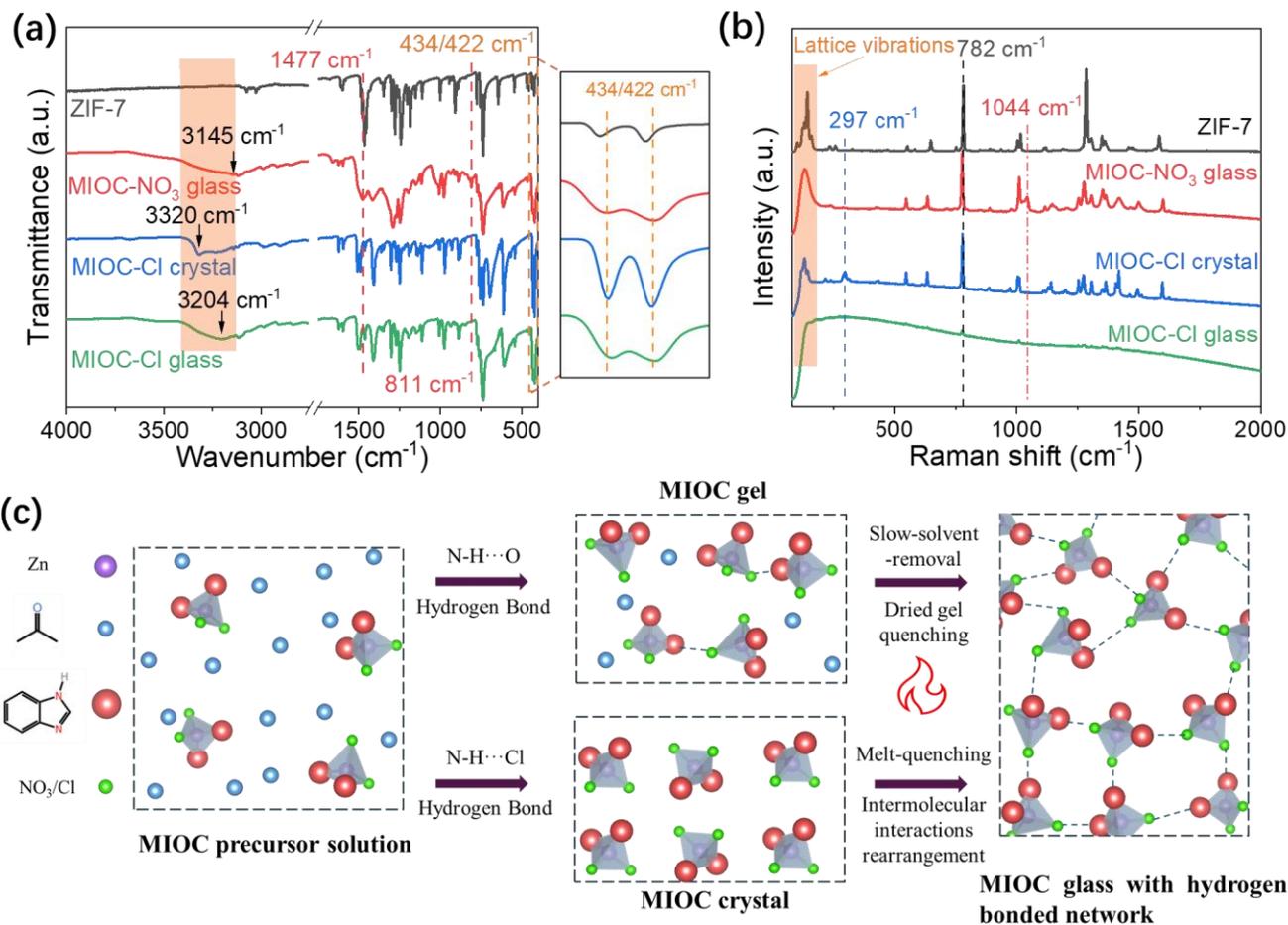

**Figure 2. (a)** FT-IR, and **(b)** Raman spectra of the synthesized MIOC-NO₃ glass, ZIF-7 crystals, MIOC-Cl crystal, and MIOC-Cl glass, respectively. **(c)** Schematic diagram of different MIOC glass formation pathways, i.e., the dried gel-quenching approach (MIOC-NO₃ glass) vs the melt-quenching method (MIOC-Cl glass).

The structural evolution during the formation of MIOC-NO₃ glass can be probed using FT-IR spectroscopy, with spectra acquired during stepwise drying process and following a high-temperature heat treatment (Figure S5). The FT-IR spectrum of the MIOC-NO₃ gel closely resembles that of the final MIOC-NO₃ glass, indicating that the fundamental $Zn(NO_3)_2HbIm_2$ structural units are formed before gelation. A peak at 1700 cm$^{-1}$, ascribed to the C=O stretching mode of free acetone, confirms the presence of free solvent within the gel [29]. This peak progressively weakens and ultimately disappears as drying proceeds, signifying the effective solvent removal. Following this, a high-temperature treatment leads to a noticeable reduction of the N-H stretching peak, suggesting that the thermal decomposition of the $Zn(NO_3)_2HbIm_2$ unit disrupts the hydrogen-bond network [16]. This disruption drives the transformation toward a disordered coordination structure.

Raman spectra (Figure 2b) provide further insight into the structural features between the crystalline and glassy states of the MIOCs, particularly through their low-frequency vibrational modes. The sharp peak below 200 cm$^{-1}$ in the ZIF and MIOC crystals is assigned to lattice vibration modes [18]. However, this peak is obscured by a broad envelope of low-frequency phonon vibration modes in the MIOC-NO$_3$ and MIOC-Cl glasses, reflecting the loss of long-range order. In addition, the spectrum of MIOC-Cl crystal exhibits a distinct peak at 297 cm$^{-1}$, attributed to Zn-Cl stretching [30], which is not observed in MIOC-Cl glass due to the strong fluorescence. Despite these differences in the metal-coordination environment, all samples share common features originating from the organic ligands. For instance, the peak at 782 cm$^{-1}$, associated with the aromatic ring in-plane bending of HbIm/bIm [31], is clearly resolved in all samples. This is notable even for the MIOC-Cl glass, whose spectrum suffers from a high fluorescence background. Additionally, the peak at 1044 cm$^{-1}$ in the MIOC-NO$_3$ glass is ascribed to the N-O stretching mode of the NO$_3$ ligand [32]. Based on the IR and Raman results, it is concluded that the strength of hydrogen bonding is a critical factor governing the formation mechanisms of both MIOC-NO$_3$ and MIOC-Cl glasses, as schematically illustrated in Figure 2c.

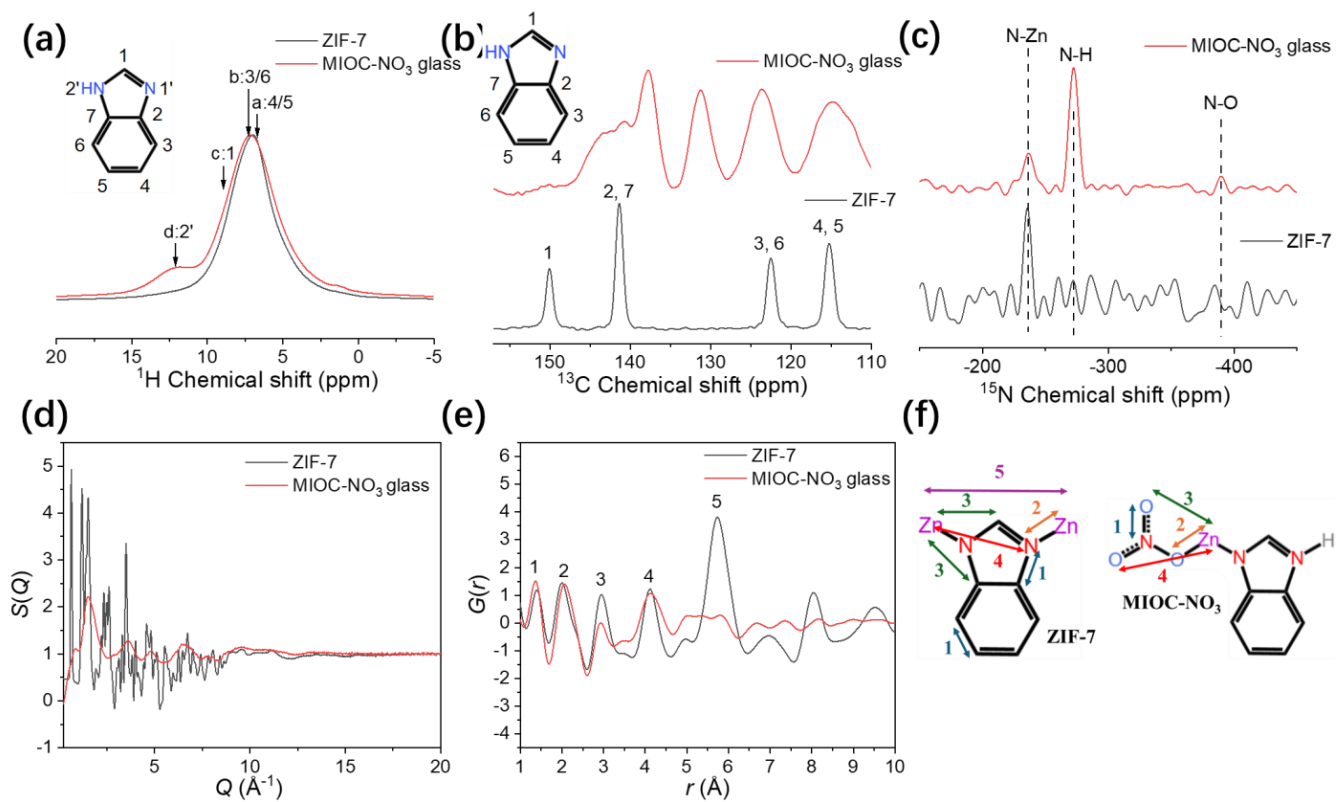

**Figure 3. (a)** Normalized $^1$H Solid-state magic-angle spinning (MAS) nuclear magnetic resonance (NMR) spectra, **(b)** Normalized $^{13}$C-$^1$H cross-polarization MAS (CPMAS) NMR spectra, **(c)** Normalized $^{15}$N CPMAS NMR spectra,

**(d)** Faber-Ziman structure factor $S(Q)$, **(e)** Reduced pair correlation functions $G(r)$ of ZIF-7 crystal and MIOC-NO$_3$ glass, respectively. **(f)** Schematic diagram of the atomic relationships in the ZIF-7 crystal and MIOC-NO$_3$ glass structure.

To probe the differences in atomic environments between non-meltable ZIF crystal and MIOC glass, $^1$H, $^{13}$C, and $^{15}$N magic-angle spinning (MAS) NMR spectra for crystalline ZIF-7 and MIOC-NO$_3$ glass are acquired (Figures 3a-c). The $^1$H NMR spectrum of crystalline ZIF-7 (Figure 3a) is characterized by a single broad resonance at ~ 7.0 ppm, which is attributed to the aromatic protons in bIm linker [21]. In contrast, the spectrum of the MIOC-NO$_3$ glass reveals a distinct profile with two separate broad resonances at 7.1 and 11.8 ppm, corresponding to aromatic protons and N-H protons of the HbIm ligand, respectively. To resolve the overlapping signals and elucidate the distinct chemical environments of H atoms, a detailed deconvolution analysis is performed on both spectra. For ZIF-7 (Figure S6), the deconvolution reveals three resonances at 6.9, 7.0, and 8.2 ppm [33], [34], respectively. Similarly, deconvolution of the MIOC-NO$_3$ glass (Figure S7) confirms the peak shift and the emergence of a new resonance at 12.2 ppm. Specifically, the signal at 6.9 ppm in ZIF-7 is assigned to the H atoms on C4 and C5 (Figure 3f), which experiences a slight shift to 6.7 ppm in MIOC-NO$_3$ glass. The resonance at 7.0 ppm, assigned to the H atoms on C3 and C6, undergoes a slight downfield shift to 7.2 ppm in the MIOC-NO$_3$ glass. More significantly, the resonance at 8.2 ppm, attributed to the H atoms on C1, experiences a pronounced downfield shift to 9.2 ppm. A new resonance appears at 12.2 ppm in the glass, which is assigned to N-bound protons (N-H) of the HbIm ligand. This feature is absent in the spectrum of the ZIF-7 crystal, confirming that all N atoms are coordinated to Zn$^{2+}$ (Zn-N) in ZIF-7. The observed new peak and these systematic chemical shift changes provide strong evidence for the formation of N-H···O hydrogen bonds between the HbIm ligands and adjacent NO$_3$ ligands in the MIOC-NO$_3$ glass. This interaction results in a deshielding effect on the HbIm protons due to the electron-withdrawing NO$_3^-$ group, which reduces the electron density around the HbIm protons. This effect is more pronounced for the H atom closer to the N-H bond.

Figure 3b shows the $^{13}$C NMR spectra of crystalline ZIF-7 and MIOC-NO$_3$ glass, which reveal profound differences in their local carbon environments. For ZIF-7, the coordination of all N atoms to Zn results in a symmetric structure, giving rise to four well-resolved aromatic carbon resonances. These signals, observed at 150, 141, 122, and 115 ppm, are assigned to C1, C2/C7, C3/C6, and C4/C5, respectively [21]. In contrast, the spectrum of the MIOC-NO$_3$ glass exhibits a profile that more closely resembles that of a monodentate HbIm ligand, indicating a lower symmetry of the hydrogen bond network. Furthermore, hydrogen bonding with adjacent nitrate anions introduces additional splitting and complexity in the resonances [35], [36], [37], which serves as an indirect spectroscopic signature of the disordered, hydrogen-bonded network. The $^{15}$N NMR spectra of ZIF-7 and MIOC-NO$_3$ glass (Figure 3c) provide evidence for the distinct chemical environments of nitrogen in the two materials. A resonance exclusive to the glass at -272 ppm is assigned to N atoms in N-H bonds [35] while the peak at -390 ppm, attributed to the nitrate (NO$_3$) nitrogen [38], is also absent in ZIF-7. Conversely, the signal at -236 ppm, corresponding to Zn-coordinated nitrogen atoms, is observed in both samples, confirming that the Zn-N coordination is preserved during the glass formation [39]. These spectral features reveal two critical roles for the incorporated NO$_3$ groups. First, they increase the asymmetry of the HbIm ligands and exert a strong electron-withdrawing effect. Second, the site-dependent effect of the added NO$_3$ on HbIm reflects a form of local order within the overall amorphous network. Specifically, the NO$_3$ anions occupy Zn coordination sites and form N-H···O hydrogen bonds with adjacent HbIm ligands, rather than existing as a free ion or a nitrated group. These $^{15}$N NMR results offer robust support for the proposed structural model of the MIOC-NO$_3$ glass, which is composed of uniformly coordinated [Zn(ligand)$_4$] tetrahedral units connected by a disordered hydrogen-bonded network involving the NO$_3^-$ anions.

To probe the short-, medium-, and long-range structure in MIOC glasses, total X-ray scattering measurements are conducted on crystalline ZIF-7 and MIOC-NO$_3$ glass. As shown in Figure 3d, the Faber-Ziman structure factor $S(Q)$ of crystalline ZIF-7 exhibits sharp diffraction peaks, characteristic of its long-range lattice order. The MIOC-NO$_3$ glass displays only broad and diffuse halos, confirming the long-range disordered structure [40]. The short-range order was elucidated by the reduced pair correlation functions, $G(r)$, (Figure 3e) of both ZIF-7 and MIOC-NO$_3$ glass, with specific atom-atom correlations illustrated in Figure 3f. Peak assignments were guided by previous reports [16], [41], [42], [43]. The first peak at 1.4 Å (Peak 1) is attributed to the aromatic C-C/C-N bonds within the benzimidazole ring, which overlaps with the N-O bonds from NO$_3$ ligands. Because the N-O bond length (~1.3 Å) is shorter than C-C/C-N (~1.4 Å), this peak shifts slightly to a lower $r$ value in the MIOC-NO$_3$ glass. The second peak at 2.2 Å (Peak 2) arises from the overlap of nearest-neighbor Zn-N (2.0 Å) and Zn-O (2.1 Å) correlations, providing direct evidence that the Zn center is coordinated by both imidazolate and nitrate ligands in MIOC-NO$_3$ glass. Peak 3 at approximately 2.95 Å is assigned to Zn correlations with its next-nearest-neighbor atoms, specifically the C atoms from the organic ligand (Zn-C) and the oxygen atoms from the NO$_3$ ligands (Zn-O). Peak 4 at approximately 4.1 Å is attributed to the combined contributions of Zn correlations with next-neighbored N (in the organic HbIm ligand) and the third-neighbored O (from NO$_3$ ligands). The superposition of these distinct atomic distances leads to the broadening of this peak in the glass structure. Peak 5 at 7.5 Å, a feature attributed to nearest-neighbor Zn-Zn correlations. In contrast to crystalline ZIF, the MIOC-$NO_3$ glass lacks this characteristic. This absence confirms the medium-range disorder in the glass network. The $G$(r) curve for MIOC-$NO_3$ glass indicates that the correlations between Zn and HbIm, i.e., the fundamental [Zn(HbIm)$_4$] tetrahedral units, are preserved, as evidenced by the persistence of key peaks associated with the C-C/C-N, Zn-N, and Zn-C peaks in ZIF-7. Furthermore, the distinct Zn-O correlation provides clear evidence for the coordination of the $NO_3$ ligand to the metal center in the glass. In contrast to this preserved short-range order, the glass lacks the characteristic Zn-Zn correlation peak at 7.5 Å (Peak 5), which is prominent in crystalline ZIF. The absence of this peak signifies a breakdown of intermolecular connectivity, leading to the loss of medium-range order.

The above-presented results reveal that the MIOC-$NO_3$ glass is fundamentally constructed by the Zn-ligand tetrahedral units interconnected by a disordered hydrogen-bonding network, with the strength of these hydrogen bonds determining the vitrification pathway. For the melt-quenching pathway (MIOC-Cl glass), medium-strength hydrogen bonds (e.g., N-H···Cl) are insufficient to direct assembly alone and instead cooperate with weaker interactions, such as weaker hydrogen bonds and π-π stacking, to form an ordered crystalline phase in solution. Upon melting, this ordered network is disrupted, allowing structural rearrangement into a disordered state. Subsequent quenching rapidly freezes the disordered network, wherein medium-strength hydrogen bonds become the dominant intermolecular force. In contrast, the direct solution route is enabled by strong hydrogen bonds (e.g., N-H···O), which effectively suppress the molecular recognition and assembly necessary for crystallization. Consequently, upon solvent removal, the system bypasses the crystalline phase to form the MIOC-$NO_3$ glass directly. A consistent feature across both pathways is the remarkable stability of the primary Zn-ligand coordination units, which remain intact throughout the structural transformations. This implies that the vitrification is driven by the formation or reorganization of the disordered network among these pre-formed Zn-ligand coordination units, rather than by the breakdown of the coordination bonds themselves.

### 3.3. Multifunctional performance characterization of MIOC glass

### 3.3.1. Gas absorption performance of MIOC-$NO_3$ glass

To evaluate the application potential of the processability of the MIOC-$NO_3$ glass, its gas adsorption and optical properties were preliminarily investigated. Notably, the MIOC-$NO_3$ glass transforms into a foam upon thermal treatment at 250 °C for 1 hour, prompting an evaluation of its adsorption capabilities in both states. The $N_2$ adsorption isotherms at 77 K reveal that both the as-synthesized glass and the thermally treated MIOC-$NO_3$ foam exhibit low porosity, with maximum $N_2$ uptakes of only 4.87 and 2.97 $cm^3 \cdot g^{-1}$, respectively (Figure 4a). This implies a predominantly non-porous or densely packed structure in both samples. However, a strikingly different behavior is observed for $CO_2$ adsorption at 195 K. While the as-synthesized MIOC-$NO_3$ glass shows a moderate capacity of 9.52 $cm^3 \cdot g^{-1}$, the thermally treated sample exhibits a nearly threefold enhancement to 27.78 $cm^3 \cdot g^{-1}$ (Figure 4b). In addition, both samples display pronounced hysteresis in the $CO_2$ gas desorption isotherms. This significant improvement in $CO_2$ adsorption, coupled with the pronounced hysteresis in the $CO_2$ desorption branch, suggested that the adsorption mechanism involves specific interactions between the $CO_2$ molecules and the glass matrix, rather than simple physisorption within rigid pores. The BET analysis reveals a substantial increase in the $CO_2$ surface area (from 10 to 41 $m^2 \cdot g^{-1}$) and total pore volume (from 0.002 to 0.014 $cm^3 \cdot g^{-1}$) of the MIOC-$NO_3$ glass upon thermal treatment. This makes the $CO_2$ adsorption capacity of the treated MIOC-$NO_3$ foam comparable to that of traditional melt-quenched ZIF glass [13], [44]. These findings, combined with TGA and FT-IR analysis, indicate that high-temperature treatment leads to ligand decomposition in the MIOC-$NO_3$ glass. This process disrupts the hydrogen-bonding network, resulting in the formation of a new, more open disordered coordination structure. Consequently, the decomposition triggers a foaming effect, which enhances the overall porosity and $CO_2$ gas adsorption capacities of the glass. However, this structural rearrangement also results in a reduction of pore size, which explains the very low uptake of $N_2$, a characteristic similar to that of ZIF glass [45].

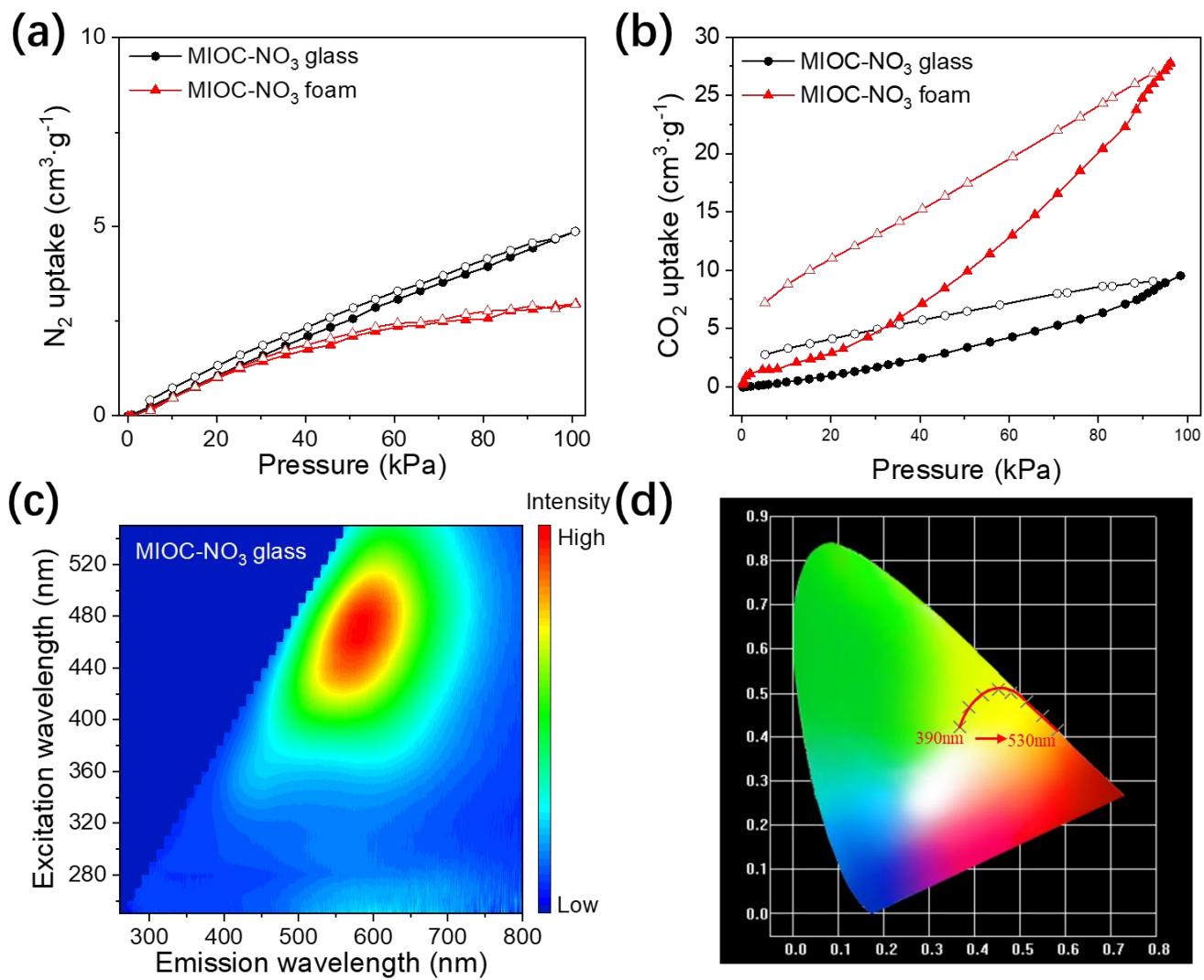

**Figure 4. (a)** $N_2$ adsorption-desorption isotherms at 77 K for both as-synthesized and the thermally treated MIOC-$NO_3$ glass. **(b)** $CO_2$ adsorption-desorption isotherms at 195 K for both as-synthesized and thermally treated MIOC-$NO_3$ glass. **(c)** Two-dimensional (excitation-emission) fluorescence spectra of MIOC-$NO_3$ glass. **(d)** CIE (Commission Internationale de l´Eclairage) chromaticity of MIOC-$NO_3$ glass.

### 3.3.2. Optical performance of MIOC-$NO_3$ glass

Beyond gas adsorption, the MIOC-$NO_3$ glass also exhibits intriguing photonic properties, making it a multifunctional material. As shown in its UV-visible absorption spectrum (Figure S8), the optical absorption edge of glass is located at 534 nm. To explore its potential fluorescence, a two-dimensional (excitation-emission) fluorescence was recorded (Figure 4c, Figure S9). Under the 470 nm laser excitation, the glass exhibits a broadband emission centered at 582 nm. Notably, the emission peak is excitation-dependent, systematically red-shifting from 540 to 616 nm as the excitation wavelength gradually increases from 380 to 550 nm, explaining the high absorption cutoff edge in the UV-visible absorption spectrum. The emission in the glass is attributed to π-π* electron transition of HbIm and electron transfer interactions with nitrate anions [10], [46]. This excitation-dependent photoluminescence is attributed to the red-edge excitation effect caused by the slow relaxation of molecules in a rigid network system. Molecules in different conformations within the glass are selectively excited by different excitation lights and emit fluorescence before the conformational transition, exhibiting a gradually red-shifted emission as the excitation light wavelength increases [47]. The corresponding CIE chromaticity is shown in Figure 4d. In contrast to the blue-to-green light emission in ZIF glasses [48], MIOC-NO$_3$ glass shows emission across the green-to-red light region, highlighting its unique potential for advanced photonic applications.

## 4. Conclusions

We report a simple low-temperature approach for directly synthesizing a stable metal inorganic-organic complex (MIOC) glass, i.e., Zn(NO$_3$)$_2$(HbIm)$_2$. Our strategy combines crystallization suppression and a step-drying process. An aprotic solvent is utilized to suppress crystallization, while a dried gel is obtained via carefully controlled slow desolation. The slow-solvent-removal process enables the formation of a robust, disordered hydrogen-bonded network around intact [Zn-N/O] tetrahedral units. Subsequent quenching of the dried gel at room temperature freezes the hydrogen-bonded network, thereby leading to glass formation.

Structural characterization revealed that MIOCs with strong hydrogen bonds tend to form disordered hydrogen-bonded networks in solution, leading to gel formation and direct vitrification upon solvent removal. MIOCs with intermediate-strength hydrogen bonds crystallize in solution but can still be transformed into hydrogen-bonded network glasses via melt-quenching. The sacrifice of hydrogen bonds ensures the preservation of the tetrahedral unit during the vitrification process. In comparison, ZIF-7, which lacks hydrogen bonds, decomposes before melting. These findings underscore the critical role of hydrogen bonding and its strength in the formation of MC glasses.

The as-synthesized MIOC glass exhibits rapid sub-$T_g$ relaxation at room temperature, accompanied by an increase in its $T_g$ from 333 to 348 K. The MIOC glass exhibits CO$_2$ adsorption capabilities, with a BET surface area of 10 m$^2$·g$^{-1}$ and total pore volume of 0.002 cm$^3$·g$^{-1}$. Furthermore, it exhibits unique excitation-dependent photoluminescence, with emission tunable from 540 to 616 nm upon excitation above 380 nm. Our new approach has a great potential to realize the scalable fabrication of functional MIOC glasses with tailored properties.

## Acknowledgments

The authors wish to thank the China Scholarship Council (grant 202208310046) for Tianzhao Xu's scholarship. Kenji Shinozaki acknowledges the proposal number 2024A1372 and Dr. Hiroki Yamada and Dr. Seiya Shimono (Japan Synchrotron Radiation Research Institute) for their help in HEXRD measurements and analysis. Yanfei Zhang acknowledges the financial support from Taishan Youth Scholar Project of Shandong Province (tsqn202103098), the National Natural Science Foundation of China (52472005), and the Project of Integrated Innovation of Education, Science and Industry of Qilu University of Technology (Shandong Academy of Sciences) (2025ZDZX10). The authors also thank the financial support from the State Key Laboratory of Silicate Materials for Architectures (SYSJJ2025-11).

## Competing interests

The authors declare that they have no known competing financial interests or personal relationships that could have appeared to influence the work reported in this paper.